\documentclass{memsupp}
\usepackage{natbib}\usepackage{txfonts}\usepackage{balance}
\usepackage{graphicx}
\usepackage[hyphens]{url}
\usepackage[a4paper,breaklinks,dvipdfm]{hyperref}
\idline{24}{138}
\begin{document}
\def\teff{$T\rm_{eff }$}
\def\kms{$\mathrm {km s}^{-1}$}
\newcommand{\cobold}{\ensuremath{\mathrm{CO}^5\mathrm{BOLD}}}
\newcommand{\linfor}{Linfor3D}
\newcommand{\xx}{\ensuremath{\mathrm{1D}_{\mathrm{LHD}}}}
\newcommand{\mD}{\ensuremath{\left\langle\mathrm{3D}\right\rangle}}

\title{
Molecular bands in extremely metal-poor  stars
}

   \subtitle{Granulation effects}

\author{
P.\,Bonifacio\inst{1} 
\and E.\, Caffau\inst{2,1}\thanks{Gliese Fellow}
\and H.-G. \, Ludwig\inst{2,1}
\and M.\, Spite\inst{1}
\and B.\, Plez\inst{3}
\and M.\, Steffen\inst{4,1}
\and \hbox{F.\, Spite\inst{1}} 
          }

  \offprints{P. Bonifacio}

\institute{
GEPI Observatoire de Paris, CNRS, Universit\'e Paris Diderot,
F-92195 Meudon Cedex, France
\and
Zentrum f\"ur Astronomie der Universit\"at Heidelberg,
Landessternwarte, K\"onigstuhl 12, 69117 Heidelberg, Germany
\and
LUPM, CNRS, UMR 5299, Universit\'e de Montpellier II, F-34095 
Montpellier Cedex 05, France
\and
Leibniz-Institut f\"ur Astrophysik Potsdam, An der Sternwarte 16, 
D-14482 Potsdam, Germany
}

\authorrunning{Bonifacio }

\titlerunning{Molecular bands in EMP stars}

\abstract{
The bands of diatomic molecules are important
abundance indicators, especially in metal-poor stars, 
where they are still measurable in metallicity regimes
where the atomic lines of their constituting
metallic elements have become vanishingly small. 
In order to use them for abundance determinations
it is imperative to understand the formation
of these bands. In this contribution we report
on our results obtained using \cobold\
hydrodynamical simulations. Some effects
that are qualitatively different from what found
in 1D computations are highlighted. 
Due to the large number of lines that form the bands, 
their spectrum synthesis is computationally challenging. 
We discuss some of the computational strategies we
employed to parallelise the computation and possible
future developments.

}
\maketitle{}

\section{Introduction}

Molecular bands of CH, CN and NH are the only available
indicators for the abundances of carbon and nitrogen
for stars with metallicity below [M/H]=--2.0. 
Sometimes also the bands of CO (IR) and C$_2$ can be used.
Oxygen can be measured for giant stars from the 
[OI] 630\,nm line, down to a metallicity of about
--3.5. At these low metallicities, 
for Turn-Off and Main Sequence stars only
the OH lines are available to get a handle on the 
oxygen abundance. 
For this reason understanding the formation of 
the molecular bands in the Extremely Metal-Poor (EMP) stars
is crucial for the understanding of the chemical evolution
of the elements C, N, and O. 
It has been known for some time that the
formation of these bands is strongly affected by granulation
effects \citep{asplund,collet,B1,B2,jonay}.
In this contribution to the 2012 \cobold\ workshop
we shall use the CIFIST grid of  \cobold\ models \citep{Ludwig}
to investigate these granulation effects and show some
striking differences found when using a \cobold\ model
compared to what is found when using a 1D model in hydrostatic equilibrium.
As 1D comparison, for each \cobold\ model
we use a  corresponding \xx\ model. LHD is a code that
computes  one dimensional, plane parallel 
lagrangian hydrodynamic model atmospheres that use
the same micro-physics and computational schemes for radiative
transfer as \cobold\ \citep{CL07}.

We shall also discuss the computational strategies that
we employed to make the problem tractable with the current
version of \linfor .

\section{Computing molecular bands with \linfor}

The current version of \linfor
\footnote{
\href{http://www.aip.de/~mst/linfor3D_main.html}{http://www.aip.de/\textasciitilde mst/linfor3D\_main.html}
}
is not optimised for handling a large number of lines. 
This poses a problem for the computations of molecular bands
that are often constituted by a large number of weak lines.
Although we know how
a considerable speed-up in the computation of synthetic spectra
could be achieved, 
mainly through pre-tabulation of all the thermodynamic quantities and 
interpolation in the tables, the changes to the code are not trivial and 
will require an extensive amount of testing. This is in the ``TO DO'' list
of the next three years. 

It is interesting to note that this is common to other
codes and hydrodynamical simulations,  and not peculiar to 
\cobold\ and \linfor . For example \citet{Frebel} devised an
ingenious way, modifying the log gf values in such a way
as to compute in 1D a ``3D corrected spectrum''. Though interesting
their method lacks generality and requires a ``modified'' set
of log gf's for each set of model parameters.

There are two main strategies  that we can use to make the computation
of molecular bands more tractable:
\begin{enumerate}
\item downselect the number of lines necessary to 
synthesise satisfactorily the required molecular band.  
\item make  the problem  embarrassingly parallel
(see Sect.\ref{emb}) ;
\end{enumerate}

The two strategies are not mutually exclusive
and, when appropriate we can use both. In addition they
will certainly be useful also when a more efficient
version of \linfor\ will become available.

\begin{figure}
\resizebox{\hsize}{!}{\includegraphics[clip=true]{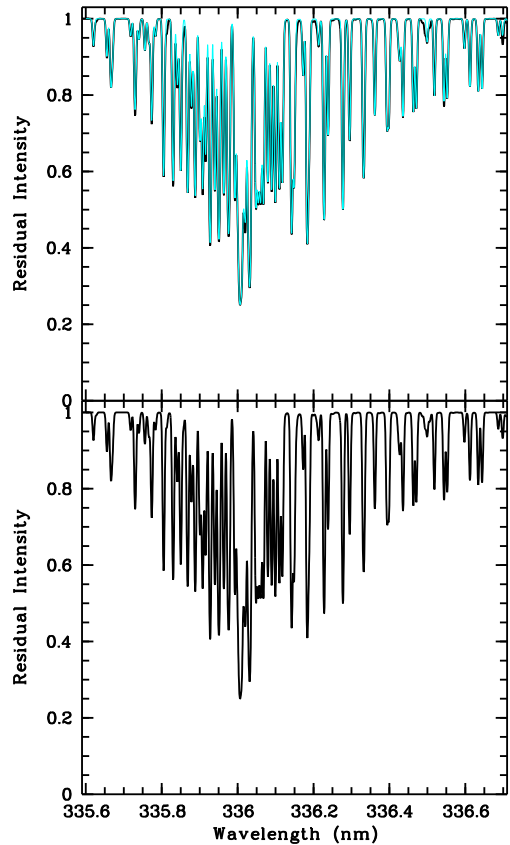}}
\caption{
The NH $A^3\Pi_i - X^3\Sigma^-$ band at 336\,nm computed
for an ATLAS 9 model 6250/3.50/--3.0 and a list of 614 lines (bottom
panel). The same spectrum compared with a spectrum
computed with a downselected line list of 138 lines (cyan, top panel).
}
\label{nh}
\end{figure}

\subsection{Downselecting the lines: the example of NH\label{down}}

For \linfor\ the time necessary to compute a weak line or a
strong line is the same, therefore it is logical to 
ask ourselves how much time it is necessary to spend
in computing very weak lines, and what is the threshold
below which we may just omit a line without {\em significantly}
changing the computed feature. 
The easiest way is to use 1D computations, that are fast, as
a guideline. We did such an exercise for the NH 
$A^3\Pi_i - X^3\Sigma^-$ band at 336\,nm using
a 1D ATLAS 9 model atmosphere \citep{K70,sbordone,K05} 
and  SYNTHE \citep{K93,sbordone,K05} spectrum synthesis. 
In Fig. \ref{nh} in the bottom panel is shown
the spectrum synthesis for a model with \teff = 6250,
log g = 3.50 and [M/H]=--3.0, using the full line list
614 NH lines as found in the Kurucz 
list\footnote{\url{http://kurucz.harvard.edu/LINELISTS/LINESMOL/nh.asc}}.
In the top panel the same spectrum is compared to another
synthetic spectrum (cyan) computed for the same model, but
with a downselected list of only 138 lines.
It is visually clear that the difference of the two computed
spectra is very small. Note that 
these spectra are not convolved with
any instrumental profile. For practical use such a spectrum
will have to be convolved with an instrumental profile
and compared to an observed spectrum of finite S/N ratio.
This example shows how we can successfully reduce the number of 
lines used in the spectrum synthesis, in this case by over
a factor of 4.
The limit of this approach is that this downselected line list
is valid only for a relatively narrow range of effective temperatures
and N abundances. Fortunately many astrophysical problems can 
be addressed by samples of stars with these characteristics.

\subsection{Making the problem embarrassingly parallel\label{emb}}

In the community of parallel computing one speaks of
a process being ``embarrassingly parallel'' when the parallelisation
is achieved so simply that it is ``embarrassingly easy'' to do.
Examples of this can be 
found in the related Wikipedia article \citep{wikipedia}.
The \linfor\ spectrum synthesis using a \cobold\ model
can be made embarrassingly parallel in several ways:
\begin{itemize}
\item Parallelise over the models. In the general
practice of abundance analysis one desires to compute
synthetic spectra for several model atmospheres, spanning
a range in atmospheric parameters. The computation
for each model is a parallel process.
\item  Parallelise over abundances. Again one needs to compute 
a profile for several abundances of the involved elements, in order 
to  determine the ``best fitting abundances''. Usually more 
than one element is involved even in the case
of hydrides. For example, all carbon-bearing molecules
are sensitive to the assumed oxygen abundance, since the chemical
equilibria have to take into account CO formation. If there are
two or more elements involved, the number of synthetic spectra
to be computed grows very rapidly.  Spectrum synthesis  for
a given set of abundances is a parallel process. 
\item Parallelise over snapshots. We select a ``small''
number of statistically independent snapshots (typically about 20)
in order to compute a time averaged synthetic spectrum. 
Spectrum synthesis for each snapshot is a parallel process.
\item Parallelise over inclinations and azimuth angles. 
Each pair of angles gives rise to a parallel process. Our
typical computation uses 3 inclinations and 4 azimuth angles, 
potentially a factor of 10 speed-up can be achieved by parallelising
over angles. As for parallelisation over snapshots, 
some overhead is expected  by book-keeping and packing the synthesis
for the different angles in a single file. 
\item Parallelise over wavelengths (wavenumbers). The process
is intrinsically parallel over wavelength, so that we may chop
up the required wavelength interval in as many wavelength
intervals as we need in order to make the problem tractable. 
In this case it is necessary to prepare many input line files
for \linfor , one for each interval. Care must be taken to  
include a few lines whose line centre is outside the
interval, but whose wings fall into the interval, on both
blue and red edges, in order to avoid unphysical ``jumps''
when the intervals are merged to produce a single  
output file. 
 
\end{itemize}

How far it is useful to push the parallelisation depends on how
many compute-cores one has available. As a rule of thumb
it is worth to parallelise up to the level that each
core is used once. 
In the computations described in this paper we have parallelised
over models, abundances and wavelengths. Parallelisation over
snapshots requires writing a book-keeping script that collects 
all the the spectra computed from individual snapshots and packs
them into a single output file, for ease of analysis.
We have not yet done this because the three adopted parallelisations
were already filling our small cluster of 56 compute cores.

\subsection{Examples of \linfor\ computations of molecular bands}

In the following we shall illustrate some of the progress 
made in this field and highlight some of the problems found and
some of the physical effects of granulation, as implied by 
our computations.

\subsubsection{The NH 336\,nm band}

\begin{figure}
\resizebox{\hsize}{!}{\includegraphics[clip=true]{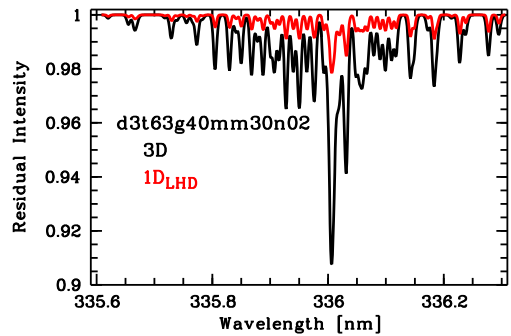}}
\caption{
The NH $A^3\Pi_i - X^3\Sigma^-$ band at 336\,nm synthesised 
for a \cobold\ model 6300/4.0/--3.0, computed with
12 opacity bins (black line) compared with the 
computation done with the corresponding \xx\ model. 
The spectra have been  
computed with the downselected line list of 138 lines 
discussed in Sect.\ref{down}.
}
\label{nh3d}
\end{figure}

Using the downselected line list discussed in 
Sect. \ref{down} and splitting the wavelength range in two
parts we  computed the full NH band. On the Paris cluster
{\tt godot} this took about 17 hours for each sub-band (computed
in parallel), for 20 snapshots, 3 inclinations and 4 azimuth 
angles. The result is shown in Fig.\,\ref{nh3d} where the
synthesis is compared with the one obtained from the
corresponding \xx\ model. As expected the band is much 
stronger in the synthesis with the \cobold\ model.   

This computing time appears acceptable, each snapshot takes
less than one hour. Parallelisation over snapshots should be
able to provide the result in about one hour and parallelisation
over angles in less than 10 minutes. These computational times
approach what is achievable in one dimensional computations. 
With two important {\em caveats}: (i) the band is split in two
sub-bands, parallelising also over 20 snapshots and 12 angle pairs
implies $2\times 20\times 12 = 480$ processes; (ii) the overheads
in accessing the snapshots and writing the results on disk should
be carefully minimised by matching the procedure to the hardware
capabilities of the system.

\begin{figure}
\resizebox{\hsize}{!}{\includegraphics[clip=true]{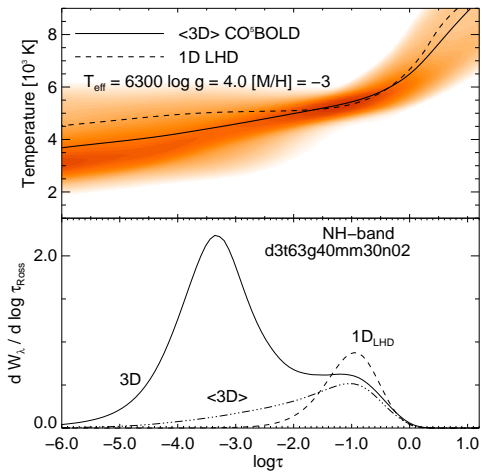}}
\caption{
Top panel: temperature
structure of the CO5BOLD model 6300/4.0/â3.0,
computed with 12 opacity bins:
the colour shading quantifies the temperature
fluctuations; the mean temperature structure, 
T$^4$ averaged over surfaces of constant
rosseland depth, is shown as a solid line,
the temperature structure of the corresponding
1D LHD model is shown as a dashed line.
Bottom panel: the  NH $A^3\Pi_i - X^3\Sigma^-$ band at 336\,nm
contribution function computed for the
CO5BOLD and \xx\ model.}  
\label{nh_contf}
\end{figure}

It is interesting to note that the \cobold\ model
is on average much cooler than the \xx\ model 
in the outer layers (see Fig.\,\ref{nh_contf}). 
Consistently the contribution function of
the \mD\ model shows an extended tail in the outer layers. 
However the cooler mean temperature is not the dominant
effect of the granulation. It is the horizontal 
temperature fluctuations  that are responsible for the large
difference between the NH band computed from the \cobold\
model and from the corresponding \xx\ model. 
This can be appreciated by the large difference between 
the contribution function computed from the \cobold\
model and that computed from the \mD\ model and from the
corresponding differences in the computed bands 
(Fig.\,\ref{nh_contf}).

We should be cautious before claiming that real stars
behave like this computation for two reasons:
\begin{enumerate}
\item the temperature structure
of  the outer layers, where our computation
puts the peak of the contribution function, could be 
radically difference in presence of a chromosphere;
\item our computation has been carried out strictly in LTE, 
in the low-density outer layers we can expect non-equilibrium
phenomena to play an important role. 
One should always remember the sobering result obtained
for lithium \citep{cayrel,asp03}, where indeed
the NLTE effects lead to a totally different
contribution function. 
\end{enumerate}

The recent observations of the 1083\,nm He I line
in metal-poor dwarf and subgiant stars by
\citet{Takeda} suggest that, contrary to common wisdom and
in spite of their old age, 
at least some of these stars possess chromospheres.
It is therefore important to pursue both observational
and theoretical studies in order to assess the role
of chromospheres in metal-poor unevolved stars. 
The presence of a chromosphere may completely change
the line formation of molecular bands and therefore the
abundances deduced from their analysis.
The \cobold\ models have provided new insight in the
solar chromosphere \citep{w04} suggesting that 
the temperature rise of the semiempirical 
chromospheric models \citep[e.g.][]{fontenla} 
likely traces only the clumps of hot material, 
in the presence of significant clumps of cold
gas. This suggests that \cobold\
extended models may  be an important theoretical
tool to investigate chromospheres in metal-poor stars.

It is equally important to pursue  theoretically the 
computations of NLTE effects both on chemical equilibria 
and on molecular line formation. 

\begin{figure*}
\resizebox{\hsize}{!}{\includegraphics[clip=true]{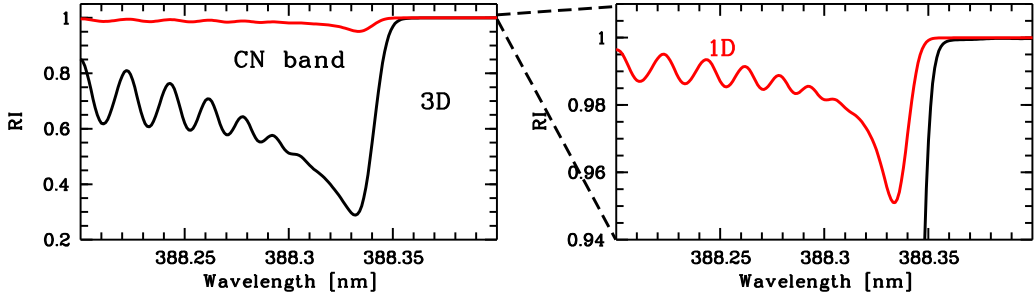}}
\caption{
The CN $B^2\Sigma-X^2\Sigma\ (0-0)$ violet 388.3\,nm band
computed with the \cobold\ model 6330/4.0/--3.0 
and the corresponding \xx\ model for a carbon-to-oxygen ratio C/O=1.
}
\label{cn}
\end{figure*}

\begin{figure*}
\resizebox{\hsize}{!}{\includegraphics[clip=true]{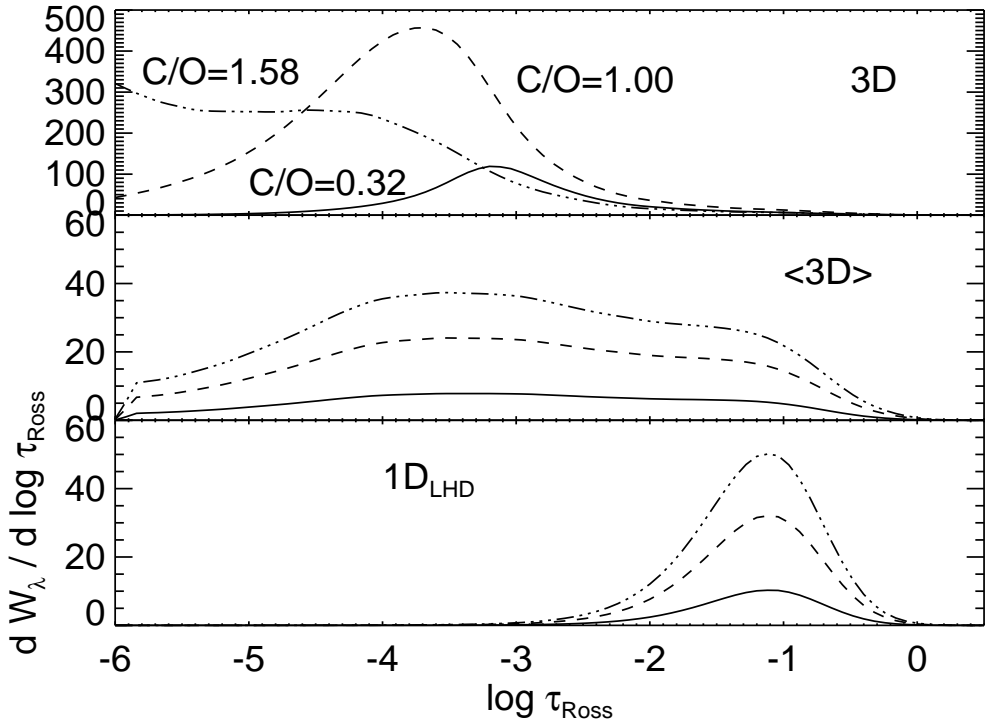}}
\caption{
The contribution function of the
CN $B^2\Sigma^+-X^2\Sigma^+ (0-0)$ violet 388.3\,nm band
computed with the \cobold\ model 6330/4.0/--3.0 (top panel),
the corresponding \mD\ model (middle panel) and \xx\ model 
(bottom panel)  for different
C/O ratios.
}
\label{cn_contf}
\end{figure*}

\subsubsection{The CN violet band at 388.3\,nm}

The computation of the CN $B^2\Sigma^+-X^2\Sigma^+ (0-0)$ violet 388.3\,nm
band
is relatively easy to tackle. The line  list computed by \citet{Plez} 
comprises 256 lines, thus we can compute it without
splitting the range in smaller domains. Note that a CH line 
falls in the band and has to taken into account.
The results are very  different 
when using a \cobold\ model or a \xx\ model. 
An example is shown in Fig.\,\ref{cn}.
The deduced nitrogen abundance, for a given 
observed spectrum,  differs by more than a factor of 100. 

This band provides a striking example of
how the physics is different in a 3D hydrodynamical
model atmosphere with respect to a corresponding 1D model
atmosphere. 
The band is, {\em a priori } expected to depend
on the oxygen abundance, however while in the
1D models, both \xx\ and \mD , this results
in a simple strengthening or weakening of the band,
while the line formation depth remains
more or less the same (see Fig.\,\ref{cn_contf}), when
using the \cobold\ hydrodynamic simulation, the line formation
depth changes in a dramatic way, moving towards outer layers
as the C/O ratio increases. 
The behaviour is indeed catastrophic
when this ratio exceeds one: the contribution function
would like to peak  at optical depths smaller than
those covered by  the model. Clearly the synthesis in this
case is not at all reliable, and should be taken only
as an indication of the trend. It is questionable whether it
is legitimate to extend the model at much lower optical
depths that log($\tau_{Ross}$)=--6, where the presence
of a chromosphere could already be felt and where the 
assumption of LTE is likely to break down. 
As pointed out in the case of the NH molecules,
both the presence of chromosphere and NLTE effects could
radically change this picture. 
It is nevertheless interesting to point out that the
effects of the cool ``clouds'', as captured by the
\cobold\ hydrodynamical simulation,  gives rise
to physical effects that are not predicted  by any
1D model.

\subsubsection{The G-band}

The G-band, the $A^2\Delta - X^2\Pi$ CH band,
is challenging from the computational point of view, 
since it is very extended, and one wants to
compute about 25\,nm of synthetic spectrum.   
Moreover the observation that the $^{12}$C/$^{13}$C
ratios in metal-poor stars range from 3 to over 100,
requires to keep track properly of both
$^{12}$CH and $^{13}$C lines. Finally in such a large
wavelength range the atomic lines cannot be neglected and
also the wing of H$\gamma$. 
This implies that to synthesise the full band on needs
to split the band in over 50 wavelength intervals, this 
calls for a large number of compute cores. 
For the time being we have just been able to synthesise the core
of the band, as shown in Fig.\, 3 of \citet{Spite13}. For this
computation we used a list of 99 lines, after a process of
downselection, similar to what we described for NH above.

\balance
\section{Lessons learnt and future developments} 

We have shown that the problem of computing
molecular bands in a \cobold\ model is
embarrassingly parallel and we have experimented
using at least three levels of parallelisation.
Full parallelisation can be achieved, and is indeed
desirable, provided enough compute cores are available. 
The actual running of these embarrassingly parallel processes
has shown some of the hardware/software limitations
that one has to take into account.  
The first limitation is that the models are stored on
a file system that is accessed through NFS by the compute cores.
We have experienced several crashes if too many processes (over 30)
are accessing the file system at the same time. 
Strategies should be sought to minimise the impact of this, 
for instance distributing the models on different file systems,
but one should consider also the use of
fast access solid state devices. 
Other remote access protocols, that may be more performing than
NFS should also be investigated.
When several tens of processes are launched and only a few crash
it turns out to be time-consuming to detect which processes have
crashed and to re-launch them. In order to perform large
scale production runs of this kind it is imperative to 
develop a software that automatically detects crashes and re-launches 
the crashed processes. One general problem of embarrassingly 
parallel processes is to have a neat book-keeping and 
easy collection of the results for further use. 
In this respect our procedures need to be greatly improved.

Computer centres with hundreds or thousands of nodes
are nowadays relatively easily accessible both at regional
and national level. To run a massive \linfor\ computation
of molecular bands we can foresee two main difficulties:
(i) time allocation committees usually frown upon embarrassingly
parallel processes and it may be difficult to convince them
that we really need a large number of processes;
(ii) most computer centres do not offer IDL.
In view of these difficulties one should consider whether 
a small (50-100 cores) dedicated cluster, that may be
bought by a given laboratory,  may not be more
effective than transferring the computation to a computer centre. 
In the next future we shall experiment with using the clusters
available at GEPI in Paris and LSW in Heidelberg to assess how
far this exercise can be pushed. In this
case the level of parallelisation is lower, for instance, no 
need to parallelise over snapshots or angles. 

We expect to be able to tackle an extensive computation
of NH, CN and CH bands throughout the whole of the
CIFIST \citep{Ludwig} grid within the next year.

\begin{acknowledgements}
PB, M. Spite and F. Spite 
acknowledge support from the Programme National
de Physique Stellaire (PNPS) and the Programme National
de Cosmologie et Galaxies (PNCG) of the Institut National de Sciences
de l'Univers of CNRS. 
EC and HGL acknowledge financial support
by the Sonderforschungsbereich SFB881 âThe Milky Way
System (DFG).

\end{acknowledgements}

\bibliographystyle{aa}

\

\end{document}